\documentstyle[aps,twocolumn,epsf,amssymb,floats]{revtex}

\newcommand{\PostScript}[3]{
  \vspace{#1cm}
\begin{center}
  \epsfysize=#2cm \leavevmode \epsfbox{#3}
\par
\end{center}
}

\begin{document}
\draft \preprint{TPR-99-22}

\title{Thermal expansion in small metal
clusters and its impact on the electric polarizability}

\author{S.\ K\"ummel$^1$, J.\ Akola$^2$, and M.\ Manninen$^2$}
\address{$^1$Institute for Theoretical Physics, University of
  Regensburg, D-93040 Regensburg, Germany\\ 
  $^2$Department of Physics, University of Jyv\"askyl\"a, P.O.\ Box
35, FIN-40351 Jyv\"askyl\"a, Finland}

\date{\today} \maketitle
\begin{abstract}

The thermal expansion coefficients of $\mathrm{Na}_{N}$ clusters with
$8 \le N \le 40$ and $\mathrm{Al}_{7}$, $\mathrm{Al}_{13}^-$ and
$\mathrm{Al}_{14}^-$ are obtained from {\it ab initio }
Born-Oppenheimer LDA molecular dynamics. Thermal expansion of small
metal clusters is considerably larger than that in the bulk and
size-dependent. We demonstrate that the average static electric dipole
polarizability of Na clusters depends linearly on the mean interatomic
distance and only to a minor extent on the detailed ionic
configuration when the overall shape of the electron density is
enforced by electronic shell effects. The polarizability is thus a
sensitive indicator for thermal expansion. We show that taking this
effect into account brings theoretical and experimental
polarizabilities into quantitative agreement.

\end{abstract}



\narrowtext \flushbottom

Since electronic shell effects were put into evidence in small
metallic systems \cite{knightp,ekardt,beck,manninen}, metal clusters
have continously attracted great interest both experimentally and
theoretically
\cite{moullet2,guan,rubio,ullrich,chelikowsky,rayane}. Besides
technological prospects, one of the driving forces for this research
has been the fundamental question of how matter develops from the atom
to systems of increasing size, and how properties change in the course
of this growing process. In some cases it has been possible to extract
detailed information from experiments done at low temperatures
\cite{expcoldopen} and the related theories \cite{koutecky96}. In many
cases, however, a deeper understanding is complicated by the finite
temperature which is present in most experiments due to the cluster
production process, see e.g.\ the discussion in \cite{brockhaus}.
Whereas a lot of theoretical information about finite temperature
effects in nonmetallic systems has been gained in the last years
\cite{thermal}, only little is known about it in metallic
clusters. Here, sodium is a particularly interesting reference system
because of its textbook metallic properties and the fact that it has
been extensively studied within the jellium model, see e.g.\
\cite{revmod} for an overview. Aluminum, on the other hand, is of
considerable technological interest. Some advances to study
temperature effects in metal clusters including the ionic degrees of
freedom were made using phenomenological molecular dynamics
\cite{bulgac}, a tight-binding hamiltonian \cite{poteau}, the
Thomas-Fermi approximation \cite{tf} or the Car-Parrinello method
\cite{roethlis}. Recently, it has also become possible to study sodium
clusters of considerable size \cite{rytkoenen} using {\it ab initio}
Born-Oppenheimer, local spin density molecular dynamics (BO-LSD-MD)
\cite{ldamd}.

In this work we report on the size dependence of a thermal 
property which is well known for bulk systems, namely the linear
thermal expansion coefficient 
\begin{equation}
\label{defbeta}
\beta=\frac{1}{l}\frac{\partial l}{\partial T}.
\end{equation}
For crystalline sodium at room temperature, it takes \cite{am} the
value $71 \times 10^{-6} K^{-1}$, for Al $23.6 \times 10^{-6}
K^{-1}$. To the present date, however, it has 
not been known how small systems are affected by thermal expansion.
At first sight, it is not even obvious how thermal expansion can
be defined in small clusters. Whereas in the bulk it is no
problem to define the length $l$ appearing in Eq.\ (\ref{defbeta}),
e.g. the lattice constant, it is less straightforward to choose a
meaningful $l$ in the case where many 
different ionic geometries must be compared to one another. For small
metal clusters, the latter situation arises because of the
many different isomers which appear at elevated temperatures.

We have calculated the thermal expansion coefficients for
$\mathrm{Na}_{8}$, $\mathrm{Na}_{10}$, $\mathrm{Na}_{12}$,
$\mathrm{Na}_{14}$, $\mathrm{Na}_{20}$ and $\mathrm{Na}_{40}$ in
BO-LSD-MD simulations. Results concerning isomerization processes in
these simulations have been presented in \cite{aisspic}, and the
BO-LSD-MD method is described in detail in Ref.\ \cite{ldamd}. A
meaningful length to be used in Eq.\ (\ref{defbeta}) if it is applied
to finite systems with similar overall deformation is the mean
interatomic distance
\begin{equation}
l_{\mathrm miad} = \frac{1}{N(N-1)}
\sum_{i,j=1}^{N} \left| {\bf R}_i - {\bf R}_j \right| ,
\end{equation}
where $ {\bf R}_i$ are the positions of the $N$ atoms in the
cluster. Obviously, $l_{\mathrm miad}$ measures the average
``extension'' of a clusters ionic structure, and we calculated it for
all configurations obtained in a BO-LSD-MD run. Two different methods
were used to calculate $\beta$. First, we discuss the heating runs, in
which the clusters were thermalized to a starting temperature
and then heated linearly with a heating rate of 5K/ps and a time step
of 5.2 fms. $l_{\mathrm miad}$ was recorded after each time step. In
this way, for $\mathrm{Na}_{8}$ the temperature range from about 50 K
to 670 K was covered, corresponding to 24140 configurations, for 
$\mathrm{Na}_{10}$ from ca. 150 K to 390
K (9260 configurations), for $\mathrm{Na}_{14}$ from ca. 50 K to 490 K
(17020 configurations), for $\mathrm{Na}_{20}$ from ca. 170 K to 380 K
(8000 configurations), and for $\mathrm{Na}_{40}$ from ca. 200 K to 400 K
(7770 configurations).

\begin{figure}[htb]
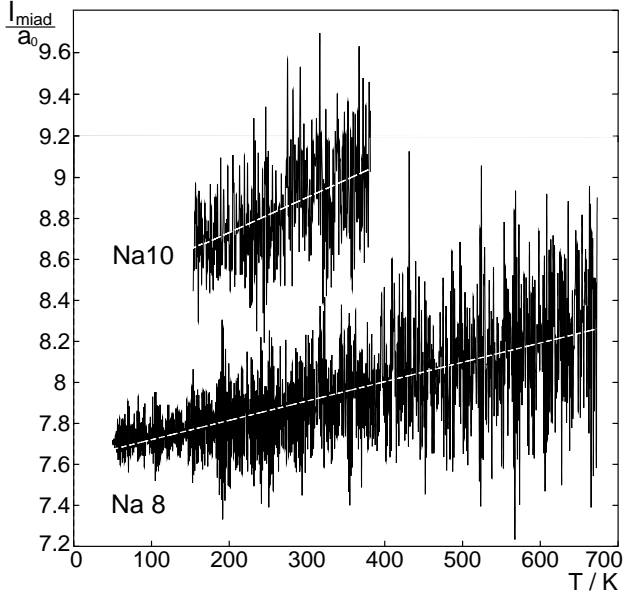

\PostScript{0}{8}{miadvt8u10.epsi}
\caption{Mean interatomic distance in $a_0$ versus temperature in K for
$\mathrm{Na}_{8}$ and $\mathrm{Na}_{10}$. The dashed lines indicate linear
fits to the complete set of data, see text for discussion. Note the
different slopes for the two clusters.}  
\label{miadvt8u10}
\end{figure}
Fig.\ \ref{miadvt8u10} shows how $l_{\mathrm miad}$ changes with
temperature for $\mathrm{Na}_{8}$ and $\mathrm{Na}_{10}$.  Both curves
show large fluctuations, as is to be expected for such small systems.
However, one clearly sees a linear rise as the general trend. We
therefore made linear fits to the data for each cluster in two ways.
The first column in the left half of table \ref{coeff} gives the
linear thermal expansion coefficients which we obtained from fitting
the data in the temperature interval between 200 K and 350 K, i.e.\
around room temperature, where bulk sodium is usually studied. In
order to allow for an estimate of the statistical quality of the fits
in view of the fluctuations, the second and third column in the left
half of Table \ref{coeff} list the ratio of the fit parameters, i.e.\
the axis interception $a$ and the slope $b$, to their standard
deviations. It becomes clear from these results that thermal expansion
in the small clusters is considerably larger than that in the
bulk. This can be understood as an effect of the increased surface to
volume ratio in the finite systems. However, the expansion coefficient
also strongly depends on the cluster size.  This can even be seen
directly from the different slopes in Fig.\ \ref{miadvt8u10}. As we
will show below, this size dependence has far reaching consequences
for the interpretation of experimental data which is usually measured
on hot clusters, as e.g.\ the static electric polarizability.

In addition to the values given in Table \ref{coeff}, we calculated the
expansion coefficient of $\mathrm{Na}_{12}$ with a different method. In two
separate runs, the cluster was thermalized to temperatures of about 200 K and
350 K, and then BO-LSD-MD was performed for 5 ps at each temperature,
i.e. without heating. From the average $l_{\mathrm miad}$ found in the two
simulations, $\beta_{\mathrm{Na}_{12}}=2.5 \, \beta_{\mathrm bulk}$ was
calculated. Thus, also the second method leads to a $\beta$ that is larger
than that of the bulk, i.e. it confirms the results of the heating runs. 

The average thermal expansion coefficient for the full temperature
range covered in each simulation is obtained from a fit to the
complete set of data, shown as a dashed line in Fig.\
\ref{miadvt8u10} for $\mathrm{Na}_{8}$ and $\mathrm{Na}_{10}$. This
average is of interest because it covers several hundred K for each
cluster in the range of temperatures which are to be expected for
clusters coming from the usual supersonic expansion sources
\cite{durgourd}. The right half of table \ref{coeff} lists these
average expansion coefficients and their statistical deviations in
the same way as before. As is to be expected, the values differ from
the previous ones for the small clusters, because the expansion
coefficient is influenced by which isomers are or become accessible at
a particular temperature, i.e.\ especially at low temperatures it is
temperature dependent. In Fig.\ \ref{miadvt8u10} one e.g.\ sees from
comparison with the average dashed line that for temperatures between
50 K and 100 K, the thermal expansion is smaller than that seen for
higher temperatures. However, once the cluster has reached a
temperature where it easily changes from one isomer to another, the
thermal expansion coefficient becomes nearly independent of the
temperature. In the case of $\mathrm{Na}_{8}$, e.g., $\beta$ changes
only by about 5 \% in the interval between 300 K and 670 K. 

Detailed previous investigations \cite{rytkoenen,aisspic} have shown
that small clusters do not show a distinct melting
transition. However, the largest cluster studied here,
$\mathrm{Na}_{40}$, shows a phase transition above 300 K
\cite{rytkoenen}. At the melting point, the octupole and hexadecupole
deformation of the electronic density sharply increase. If $l_{\mathrm
miad}$ is a relevant indicator for structural changes, then melting
should also be detectable from it. Indeed we find a noticeable
increase in $l_{\mathrm miad}$ at 300 K, and similar fluctuation
patterns as in the multipole moments. In our simulation, we could only
determine the expansion coefficient for the solid phase, and it is
given in the right half of table \ref{coeff}.
\begin{table}[tb]
\begin{tabular}{|c||c|c|c||c|c|c|}
 & $\beta / \beta_{\mathrm bulk}$ & $\sigma (a)/a$ & $\sigma (b)/b$ &
   $\beta / \beta_{\mathrm bulk}$ & $\sigma (a)/a$ & $\sigma (b)/b$ 
\\ \hline
$\mathrm{Na}_{8} $ & 2.4 & 0.001 & 0.04  & 1.7 & $<$ 0.001 & 0.01
\\ \hline                                     
$\mathrm{Na}_{10}$ & 3.6 & 0.002 & 0.03  & 2.8 & 0.001  & 0.02 
\\ \hline                                     
$\mathrm{Na}_{14}$ & 1.2 & 0.002 & 0.07  & 1.7 & $<$ 0.001 & 0.01 
\\ \hline                                     
$\mathrm{Na}_{20}$ & 1.9 & 0.001 & 0.03  & 1.9 & 0.001 & 0.01 
\\ \hline                                     
$\mathrm{Na}_{40}$ &  -  &-      &-      & 1.2 & 0.001  & 0.04 
\end{tabular}
\caption{Left half, first column: Linear thermal expansion coefficient
of small Na clusters in the temperature interval between 200 and 350 K, given
in terms of the bulk value $71 \times 10^{-6} K^{-1}$. Columns two and three
give the ratio of the axis interception $a$ and the slope $b$ to their standard
deviations as obtained from the fits.  
Right half: Expansion coefficient averaged over 50-670 K for $\mathrm{Na}_{8}$,
150-390 K for $\mathrm{Na}_{10}$, 50-490 K for $\mathrm{Na}_{14}$, 150-460 K
for $\mathrm{Na}_{20}$, and 200-300 K for $\mathrm{Na}_{40}$.
See text for discussion.} 
\label{coeff}
\end{table}

As seen in Fig.\ \ref{miadvt8u10}, $\mathrm{Na}_{8}$ shows thermal
expansion already at 50 K. This raises the question at which
temperature the expansion actually starts, i.e.\ where anharmonic
effects in the ionic oscillations will start to become important. In
this context we note that one can compare the $l_{\mathrm miad}$ at
T=0 K found by extrapolation from the heating data to the $l_{\mathrm
miad}$ which is actually found for the ground state structure at T=0
K. We have done this for $\mathrm{Na}_{8}$, $\mathrm{Na}_{10}$ and
$\mathrm{Na}_{14}$, where the ground state structures are well
established. In all cases, the differences between the two values were
less than 1\%. This indicates that the anharmonic effects for Na
clusters are important down to very low temperatures. Furthermore, the
anharmonicities should also be observable in the heat capacities
\cite{rytkoenen}, where they will lead to deviations from
Dulong-Petit's law. We have checked this and indeed found deviations
between 8 \% ($\mathrm{Na}_{20}$) and 19 \% ($\mathrm{Na}_{8}$) from
the Dulong-Petit value.

As an example for the considerable influence of thermal expansion on
measurable physical properties we discuss the average static
electric dipole polarizability $\alpha$, which is defined as the trace of the
polarizability tensor. It was one of the first observables from which the
existence of electronic shell effects in metal clusters was deduced
\cite{knightp}, and it has been measured for clusters of various sizes and
materials \cite{rayane}. For Na clusters with up to eight atoms, the
polarizability was also calculated in different approaches
\cite{moullet2,guan,rubio,chelikowsky,rayane}. These calculations 
qualitatively reproduce the experimentaly observed trends, but they all
underestimate the measured value. We show that this discrepancy
is to a large part due to the fact that the calculations were done for T=0,
whereas the measurement is done on clusters having temperatures of about 400
to 600 K \cite{durgourd}. 

\begin{figure}[htb]
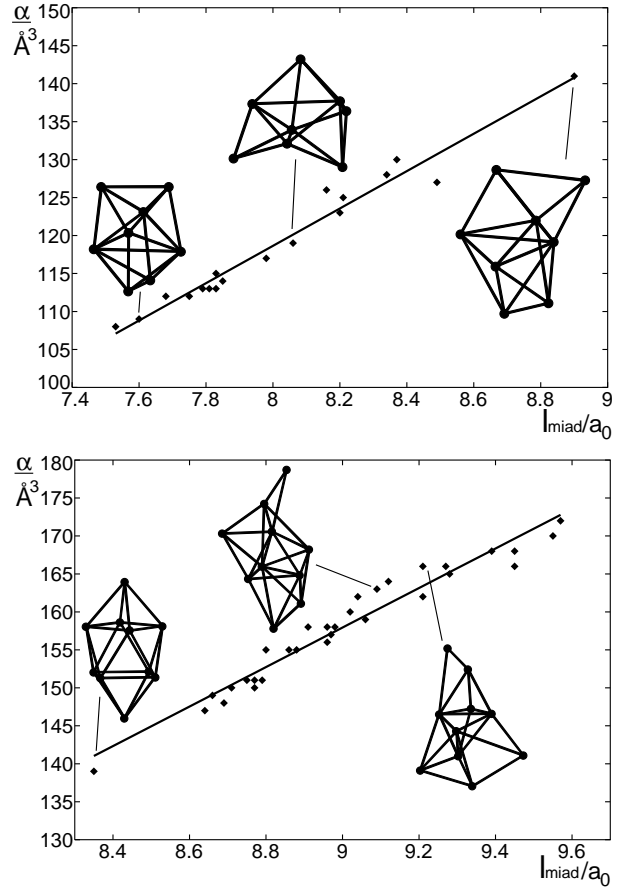

\PostScript{0}{12}{alvmiad8u10.epsi}
\caption{Static electric dipole polarizability
versus mean interatomic distance for different isomers of $\mathrm{Na}_{8}$
(upper) and $\mathrm{Na}_{10}$ (lower). Three examples of different 
geometries are shown as insets for both sizes.} 
\label{alvmiad8u10}
\end{figure}

For various, different isomers obtained in our heating runs for
$\mathrm{Na}_{8}$ and $\mathrm{Na}_{10}$, we have calculated the
polarizability from the derivative of the induced dipole moment with
respect to the electric field (finite field method). Since highly
unsymmetric isomers from the high temperature part of the simulations
were taken into account, the full tensor was computed by numerically
applying the dipole field in the different directions in seperate
calculations. We have checked that the used field strength of $5
\times 10^{-5} e/a_0^2$ is large enough to give a numerically stable
signal and small enough to be in the regime of linear response. In
Fig.\ \ref{alvmiad8u10} we have plotted the thus obtained
polarizabilities versus $l_{\mathrm miad}$, and show three instances
of ionic geometries for each cluster that demonstrate how different
the structures actually are. Nevertheless, within a few percent the
polarizabilities are on a straight line. This shows that the average
polarizability depends mainly and strongly on the mean interatomic
distance, and only to a minor extent on details in the ionic
configurations. Of course, the situation might be more complicated for
clusters where the overal shape, i.e.\ the lowest terms in the
multipole expansion of the valence electron density, is not stabilized
by electronic shell effects. For the present clusters, however, the
deformation induced by the electronic shell effects persists even at
elevated temperatures. That $\alpha$ is less sensitive to the detailed
ionic configuration than, e.g., the photoabsorption spectrum, is
understandable because it is an average quantity.

The dependence of the polarizability on the mean interatomic distance has the
consequence that $\alpha$ also strongly depends on the temperature. From
Fig.\ \ref{alvmiad8u10} one deduces that an average bondlength increase of 1
$a_0$ in $\mathrm{Na}_{8}$ and $\mathrm{Na}_{10}$ leads to an increase in
the polarizability of about 25 $\AA^3$. Thus, neglection of the
thermal expansion in T=0 calculations leads to polarizabilities which
are smaller than the ones measured on clusters coming from supersonic
expansion sources \cite{knightp,rayane}. Of course, also
underestimations of the   
cluster bond lengths that are due to other reasons will directly appear in the
polarizability. With the Troullier-Martins pseudopotential, e.g.\, the
BO-LSD-MD underestimates the dimer bond length by 4.5\%, and it is to be
expected that the situation is similar for the bond lengths of larger
clusters. Taking this into account, one can proceed to calculate the
polarizability for clusters with a temperature corresponding to the
experimental one of about 500 K \cite{durgourd}. In the
experiments the clusters are spending about $10^{-4}$s in the deflecting field
from which the polarizability is deduced, i.e.\ the experimental timescale is
orders of magnitude larger than the timescale of the fluctuations in
the mean interatomic distance (see Fig.\ \ref{miadvt8u10}). Thus, the
fluctuations will be averaged over and can be neglected. From the average
expansion coefficients we obtain a bond length increase
of 0.48 $a_0$ for $\mathrm{Na}_{8}$ and 0.87 $a_0$ for $\mathrm{Na}_{10}$ at
500 K, which in turn leads to an increase in the polarizability of 12 $\AA^3$
and 23 $\AA^3$, respectively. The resulting polarizabilities of 130 $\AA^3$ for
$\mathrm{Na}_{8}$ and 172 $\AA^3$ for $\mathrm{Na}_{10}$ compare favourably
with the experimental values 134$\pm$16$\AA^3$ and 190$\pm$20$\AA^3$
\cite{knightp,rayane}. For all other cluster sizes, the two
experiments \cite{knightp,rayane} give different values for the
polarizability. From the present work it becomes clear that
differences in the experimental temperatures might be the reason for
the discrepancies. Therefore, an accurate measurement of the clusters'
temperatures is necessary before further quantitative comparisons
can be made. However, a detailed comparison to both
experiments showed that the theoretical T=0 polarizability of all isomers
underestimates both experimental results \cite{tobepu}. Thus, the
increase in $\alpha$ that is brought about by thermal expansion will lead to 
better agreement between theory and experiment for all cluster sizes.

Thermal expansion is also observed in aluminum clusters. For
$\mathrm{Al}_{7}$ we performed 5 ps of BO-LSD-MD at each of the
fixed temperatures 100 K, 300 K,
500 K and 600 K, for $\mathrm{Al}_{13}^-$ at 260 K, 570 K and 930 K,
and for $\mathrm{Al}_{14}^-$ at 200 K, 570 K and 900 K, in analogy to
the procedure for $\mathrm{Na}_{12}$. From the
average $l_{\mathrm miad}$ at each temperature, we calculated the
expansion coefficients 
$\beta_{\mathrm Al_7}=1.3 \, \beta_{\mathrm bulk}$,
$\beta_{\mathrm Al_{13}^-}=1.4 \, \beta_{\mathrm bulk}$,
$\beta_{\mathrm Al_{14}^-}=1.4 \, \beta_{\mathrm bulk}$.
It should be noted that with $\mathrm Al_{13}^-$ we have chosen 
an electronically as well as geometrically magic cluster \cite{akola}, 
i.e.\ a particularly rigid one, and the fact that it also shows 
a larger expansion coefficient than the bulk is further evidence 
for the conclusion that the increased expansion coefficient is 
indeed a finite size effects. A noteworthy difference between 
Al and Na is seen in the temperatures where the expansion sets
in. Whereas for Na this temperature is below 50 K, we observe
that $\mathrm Al_{13}^-$ and $\mathrm Al_{14}^-$ show no
expansion below 300 K.

In summary, we have calculated thermal expansion coefficients for
small metal cluster and demonstrated that thermal expansion in these
systems is larger than that in the bulk. For the case of sodium, the
dependence of the expansion coefficient is not monotonous according to
the cluster size. We showed that the average static electric dipole
polarizability of clusters whose overall shape is fixed by electronic
shell effects depends linearly on the mean interatomic distance. Thus,
thermal expansion increases the static electric polarizability, and we
demonstrated that taking this effect into account brings the
theoretical values in a close agreement with the experimental ones.

We thank M.\ Brack and A. Rytk\"onen for clarifying discussions.
J.A. acknowledges support by the V\"ais\"al\"a Foundation, S.K. by the
Deutsche Forschungsgemeinschaft, and all authors by the Academy of
Finland.

\end{document}